\documentclass[aps,epsf,rotate,preprint]{revtex4}
\usepackage{epsfig}
\usepackage{amssymb}

\begin{document}

\title{Anomalous Transport in Complex Networks}

\author{
Eduardo L\'{o}pez$^1$, Sergey V. Buldyrev$^1$, Shlomo Havlin$^{1,2}$, 
and H. Eugene Stanley$^1$}

\address{$^1$Center for Polymer Studies, Boston University, Boston, MA
02215, USA\\ 
$^2$Minerva Center \& Department of Physics, Bar-Ilan University, Ramat
Gan, Israel}

\date{last revised: 30 November 2004 3:30PM}

\begin{abstract}

To study transport properties of complex networks, we
analyze the equivalent conductance $G$ between two arbitrarily chosen
nodes of random scale-free networks with degree distribution $P(k)\sim
k^{-\lambda}$ in which each link has the same unit resistance. We
predict a broad range of values of $G$, with a power-law tail
distribution $\Phi_{\rm SF}(G)\sim G^{-g_G}$, where $g_G=2\lambda -1$, and
confirm our predictions by simulations. The power-law tail in
$\Phi_{\rm SF}(G)$ leads to large values of $G$, thereby significantly
improving the transport in scale-free networks, compared to
Erd\H{o}s-R\'{e}nyi random graphs where the tail of the conductivity
distribution decays exponentially. 
Based on a simple physical ``transport backbone'' 
picture we show that the conductances are
well approximated by $ck_Ak_B/(k_A+k_B)$ for any pair of nodes $A$ and $B$
with degrees $k_A$ and $k_B$. Thus, a single parameter $c$ characterizes 
transport on scale-free networks.

\end{abstract}

\maketitle

Recent research on the topic of complex networks is leading to a better
understanding of many real world social, technological, and natural
systems ranging from the World Wide Web and the Internet to cellular
networks and sexual-partner networks \cite{rev-Albert}. One type of
network topology that appears in many real world systems is the
scale-free network \cite{scale-Barabasi}, characterized by a scale-free
degree distribution:
\begin{equation}
P(k)\sim k^{-\lambda},\qquad k_{\rm min}\le k\le k_{\rm max},
\label{degree}
\end{equation}
where $k$, the degree, is the number of links attached to a node. The
distribution has two cutoff values for $k$: $k_{\rm min}$, which
represents the minimum allowed value of $k$ on the graph ($k_{\rm min}=2$ here), 
and $k_{\rm max}\equiv k_{\rm min}N^{1/(\lambda-1)}$, which is the typical maximum 
degree of a network with $N$ nodes \cite{Cohen,netwcomment}.
The scale-free feature allows a network to have some nodes
with a large number of links (``hubs''), unlike the case for the classic
Erd\H{o}s-R\'{e}nyi model of random networks \cite{ER,Bollobas}.

Here we show that for scale-free networks with $\lambda\ge 2$,
transport properties characterized by conductance display a power-law
tail distribution that is related to the degree distribution $P(k)$. 
We find that this power-law tail represents pairs of nodes of high degree
which have high conductance. Thus, transport in scale-free networks is better 
than in Erd\H{o}s-R\'{e}nyi random networks.
Also, we present a simple physical picture of transport in complex networks 
and test it with our data.

The classic random graphs of Erd\H{o}s and R\'{e}nyi \cite{ER,Bollobas} have a
Poisson degree distribution, in contrast to the power-law distribution
of the scale-free case. Due to the exponential decay of the degree
distribution, the Erd\H{o}s-R\'{e}nyi networks lack hubs and their
properties, including transport, are controlled solely by the 
average degree 
$\bar{k}\equiv\sum_{i=k_{\rm min}}^{k_{\rm max}} iP(i)$
\cite{Bollobas,Grimmett-Kesten}.

Most of the work done so far regarding complex networks has concentrated
on static topological properties or on models for their growth
\cite{rev-Albert,Cohen,dyn-network,Toroczkai}. 
Transport features have not been extensively studied with the 
exception of random walks on complex networks \cite{Noh,Sood,Gallos}, despite 
the fact that transport properties contain information about network function  
\cite{dyn-topology}.
Here, we study
the electrical conductance $G$ between two nodes $A$ and $B$ of
Erd\H{o}s-R\'{e}nyi and scale-free networks when a potential difference
is imposed between them. 
We assume that
all the links have equal resistances of unit value \cite{comm-struc}.

To construct an
Erd\H{o}s-R\'{e}nyi network, we begin with a fully connected graph,
and randomly remove $1-\bar{k}/(N-1)$ out of the $N(N-1)/2$ links
between the $N$ nodes.
To generate a scale-free network with $N$ nodes, we use the 
Molloy-Reed algorithm \cite{Molloy-Reed}, which allows for the
construction of random networks with arbitrary degree distribution.  
We generate $k_i$ copies of each node $i$, 
where the probability of having $k_i$ 
satisfies Eq.~(\ref{degree}). These copies of the nodes are
then randomly paired in order to construct the network, making sure that
two previously-linked nodes are not connected again, and also excluding
links of a node to itself \cite{fn1}.  

The conductance $G$ of the network between two nodes
$A$ and $B$ is calculated using the Kirchhoff method \cite{Kirchhoff},
where entering and exiting potentials are fixed to $V_A=1$ and $V_B=0$.
We solve a set of linear equations to determine the potentials
$V_i$ of all nodes of the network. Finally, the total current $I\equiv
G$ entering at node $A$ and exiting at node $B$ is computed by adding
the outgoing currents from $A$ to its nearest neighbors through
$\sum_{j}(V_A-V_j)$, where $j$ runs over the neighbors of $A$.

\begin{figure}[t]
\epsfig{file=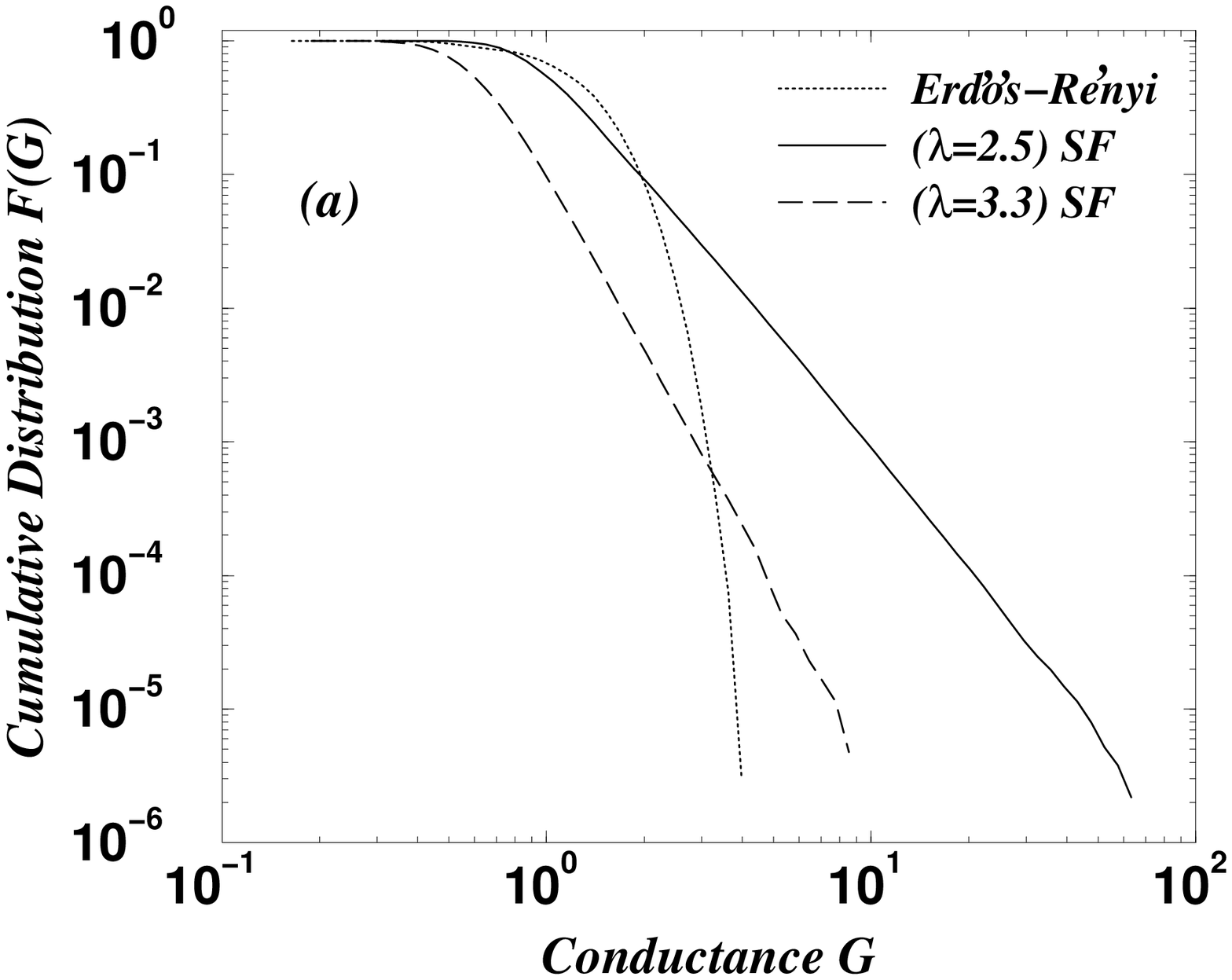,width=3in,height=2.5in}
\epsfig{file=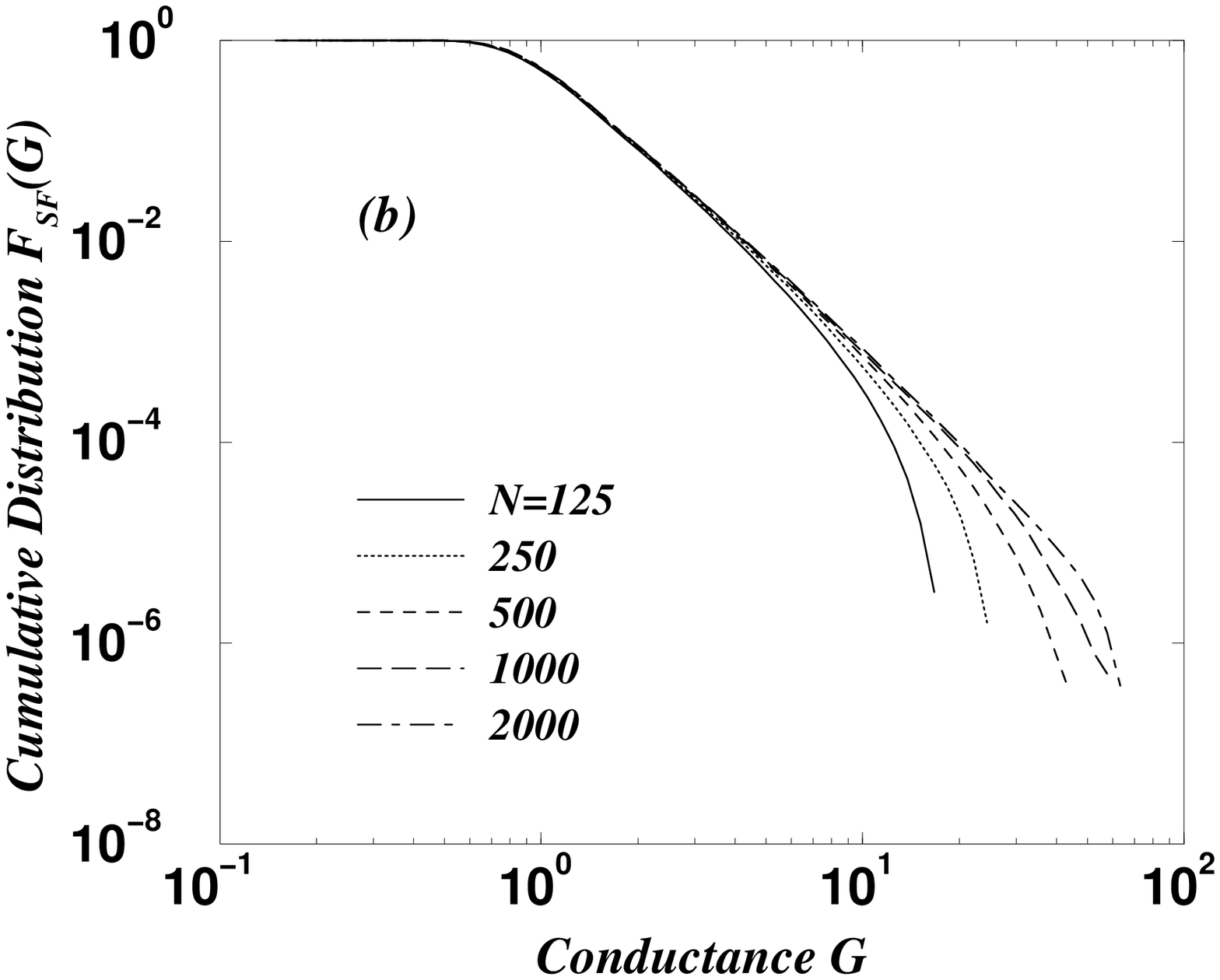,width=3in,height=2.5in}
\caption{(a) Comparison for networks with $N=8000$ nodes between the
  cumulative distribution functions for the Erd\H{o}s-R\'{e}nyi and the
  scale-free cases (with $\lambda=2.5$ and 3.3). Each curve represents
  the cumulative distribution $F(G)$ vs. $G$. The simulations have at
  least $10^6$ realizations. (b) Effect of system size on $F_{\rm
  SF}(G)$ vs. $G$ for the case $\lambda=2.5$. The cutoff value of the
  maximum conductance $G_{\rm max}$ progressively increases as $N$
  increases.  }
\label{FG_lamb2.5-3.3-ER_N8000}
\end{figure}

First, we analyze the pdf
$\Phi(G)dG$ that two nodes on the network have conductance
between $G$ and $G+dG$. 
To this end, we introduce
the cumulative distribution $F(G)\equiv\int_{G}^\infty
\Phi(G')dG'$, shown in Fig.~\ref{FG_lamb2.5-3.3-ER_N8000}(a)
for the Erd\H{o}s-R\'{e}nyi and
scale-free ($\lambda=2.5$ and $\lambda=3.3$, with $k_{\rm min}=2$) cases.
We use the notation $\Phi_{\rm SF}(G)$ and $F_{\rm SF}(G)$ for scale-free, 
and $\Phi_{\rm ER}(G)$ and $F_{\rm ER}(G)$ for Erd\H{o}s-R\'{e}nyi. 
The function $F_{\rm
SF}(G)$ for both $\lambda=2.5$ and 3.3 exhibits a tail region 
well fit by the power law 
\begin{equation}
F_{\rm SF}(G)\sim G^{-(g_G-1)}, 
\end{equation}
and the exponent $(g_G-1)$ increases with $\lambda$.  In contrast, 
$F_{\rm ER}(G)$ decreases exponentially with $G$.

Increasing $N$ does not significantly change $F_{\rm SF}(G)$
(Fig.~\ref{FG_lamb2.5-3.3-ER_N8000}(b)) except for an increase in the
upper cutoff $G_{\rm max}$, where $G_{\rm max}$ is the typical maximum
conductance, corresponding to the value of $G$ at which $\Phi_{\rm
SF}(G)$ crosses over from a power law to a faster decay. We observe no
change of the exponent $g_G$ with $N$. The increase of $G_{\rm max}$
with $N$ implies that the average conductance $\bar{G}\equiv\int
G\Phi(G) dG$ also increases slightly \cite{Gavg}.

We next study the origin of the large values of $G$ in scale-free
networks and obtain an analytical relation between $\lambda$ and $g_G$.
Larger values of $G$ require the presence of many parallel paths, which
we hypothesize arise from the high degree nodes.  Thus, we expect that
if either of the degrees $k_A$ or $k_B$ of the entering and exiting
nodes is small, the conductance $G$ between $A$ and $B$ is small since
there are at most $k$ different parallel branches coming out of a node
with degree $k$.  Thus, a small value of $k$ implies a small number of
possible parallel branches, and therefore a small value of $G$.  To
observe large $G$ values, it is therefore necessary that both $k_A$ and
$k_B$ be large.

\begin{figure}[t]
\epsfig{file=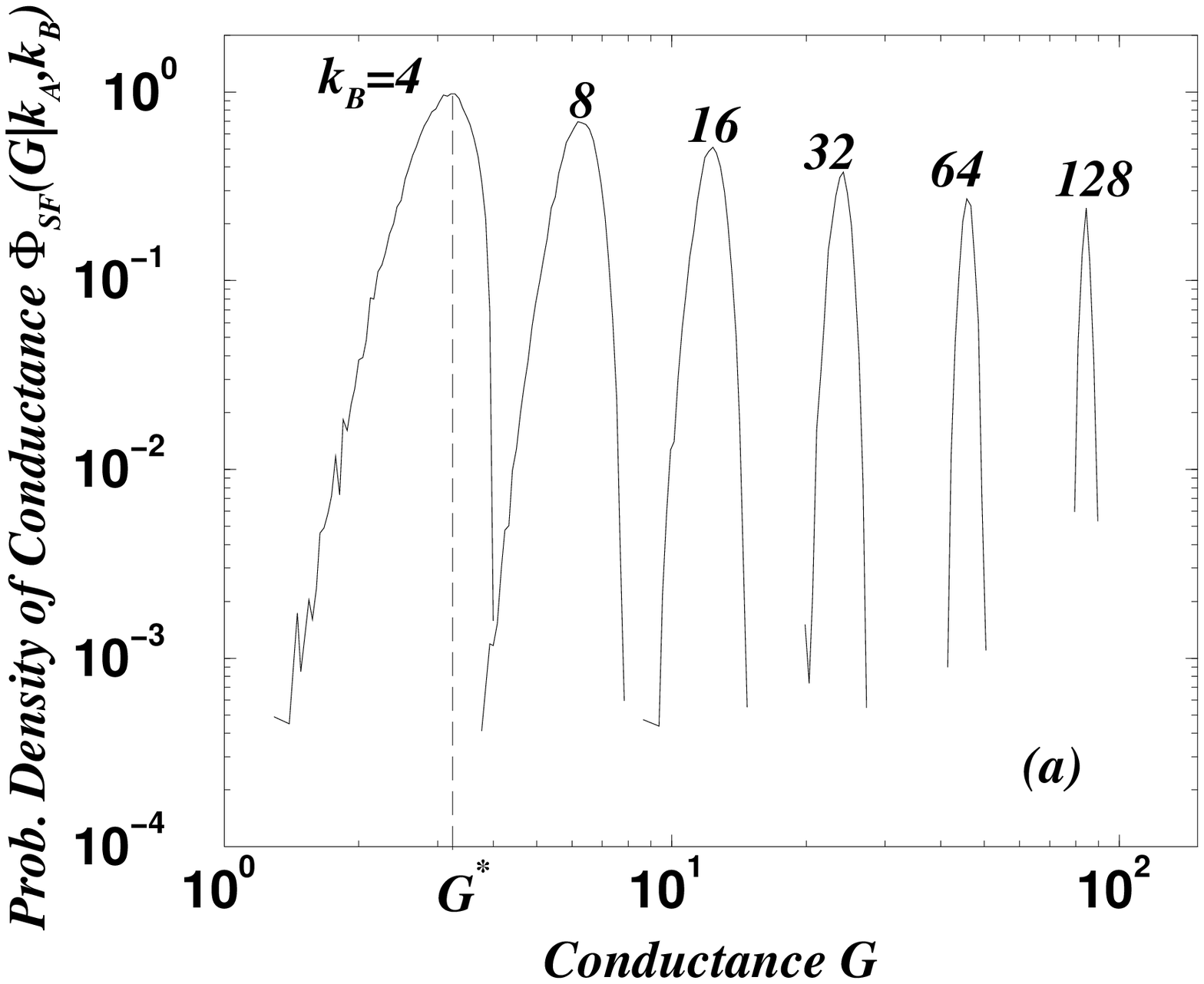,width=3in,height=2.7in}
\epsfig{file=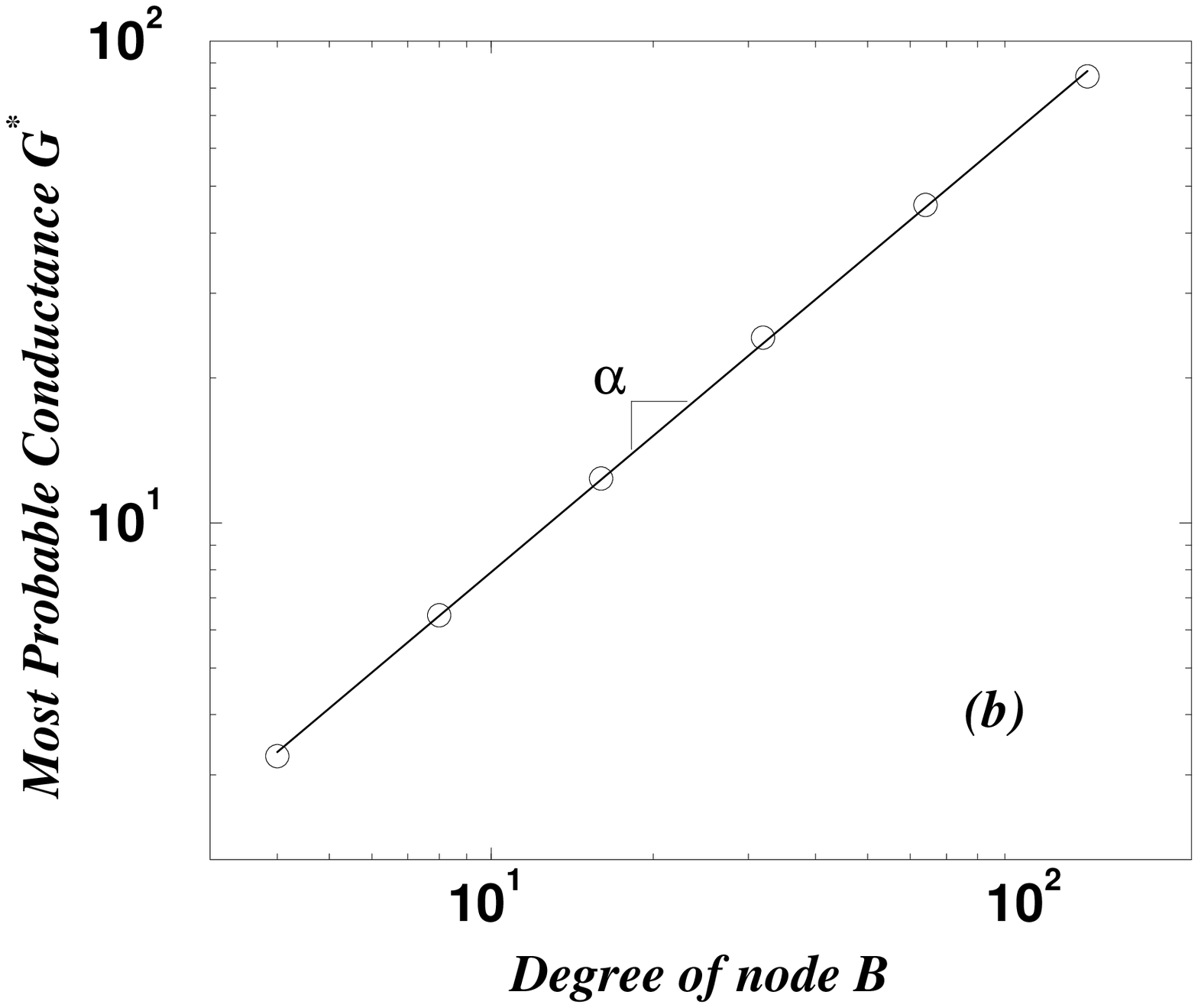,width=3in,height=2.5in}
\epsfig{file=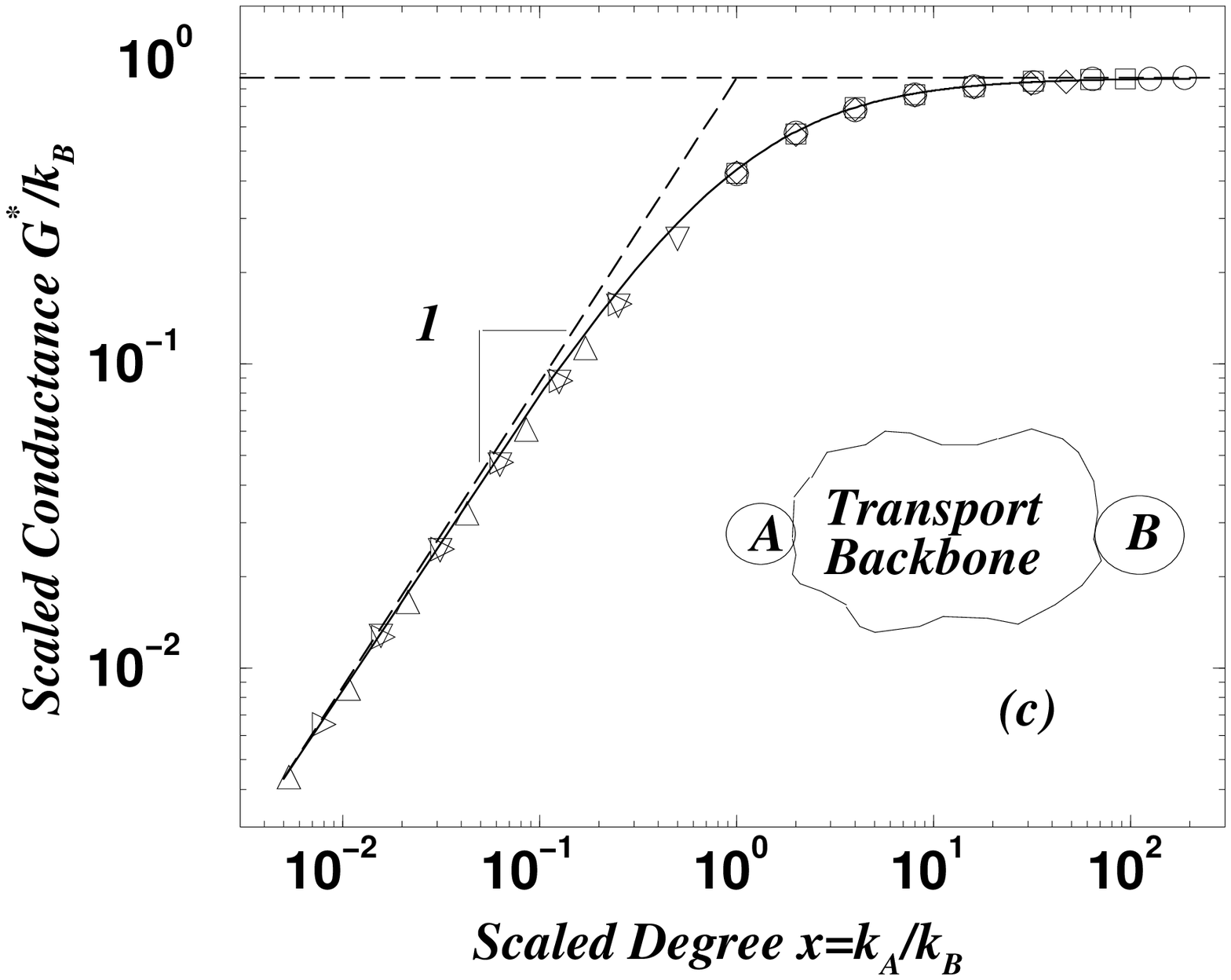,width=3in,height=2.5in}
\caption{(a) Probability density function $\Phi_{\rm SF}(G|k_A,k_B)$
     vs. $G$ for $N=8000$, $\lambda=2.5$ and $k_A=750$ ($k_A$ is close
     to the typical maximum degree $k_{\rm max}=800$ for $N=8000$). (b)
     Most probable values of $G^*$, estimated from the maxima of the
     distributions in Fig.~\ref{PGka750_kb4-128_N8000_ab7_lamb2.5}(a),
     as a function of the degree $k_B$. The data support a power law
     behavior $G^*\sim k_B^{\alpha}$ with exponent $\alpha=0.96\pm
     0.05$.  (c) Scaled most probable conductance $G^*/k_B$ vs. scaled
     degree $x\equiv k_A/k_B$ for system size $N=8000$ and
     $\lambda=2.5$, for several values of $k_A$ and $k_B$:
     $\square$ ($k_A=8$, $8<k_B<750$),
     $\diamondsuit$ ($k_A=16$, $16<k_B<750$), 
     $\bigtriangleup$ ($k_A=750$, $4<k_B<128$), 
     $\bigcirc$ ($k_B=4$, $4<k_A<750$), 
     $\bigtriangledown$ ($k_B=256$, $256<k_A<750$), and
     $\triangleright$ ($k_B=500$, $4<k_A<128$).
     The solid line is the
     predicted function $G^*/k_B=cx/(1+x)$ obtained from
     Eq.~(\ref{e4b}).  This plot shows the rapid approach of the
     scaling function $f(x)$ of Eq.~(\ref{e4b}) from a linear behavior 
     to the constant $c$ (here $c=0.87\pm0.02$, horizontal dashed line). 
     The inset shows a schematic of the ``transport
     backbone'' picture, where the circles labeled $A$ and $B$ denote the
     nodes $A$ and $B$ and their associated links which do not belong to
     the ``transport backbone''.}
\label{PGka750_kb4-128_N8000_ab7_lamb2.5}
\end{figure}

We test this hypothesis by large scale computer simulations of
the conditional pdf
$\Phi_{\rm SF}(G|k_A,k_B)$ for specific values of the entering and
exiting node degrees $k_A$ and $k_B$.  Consider first the case $k_B\ll
k_A$, and the effect of increasing $k_B$, with $k_A$ fixed.  We find
that $\Phi_{\rm SF}(G|k_A,k_B)$ is narrowly peaked
(Fig.~\ref{PGka750_kb4-128_N8000_ab7_lamb2.5}(a)) so that it is well
characterized by $G^*$, the value of $G$ when $\Phi_{\rm SF}$ is a maximum.
Further, for increasing values of $k_B$, we find
[Fig.~\ref{PGka750_kb4-128_N8000_ab7_lamb2.5}(b)] $G^*$ increases as
$G^*\sim k_B^{\alpha}$, with $\alpha=0.96\pm 0.05$ consistent with the
possibility that as $N\rightarrow\infty$, $\alpha=1$ which we assume
henceforth.

For the case of $k_B{>\atop\sim} k_A$, $G^*$ increases less fast than
$k_B$, as can be seen in Fig.~\ref{PGka750_kb4-128_N8000_ab7_lamb2.5}(c)
where we plot $G^*/k_B$ against the scaled degree $x\equiv k_A/k_B$. The
collapse of $G^*/k_B$ for different values of
$k_A$ and $k_B$ indicates that $G^*$ scales as
\begin{equation}
G^*\sim k_Bf\left(\frac{k_A}{k_B}\right).
\label{G_max_scaled}
\end{equation}
The behavior of the scaling function $f(x)$ can be interpreted using the
following simplified ``transport backbone'' picture
[Fig.~\ref{PGka750_kb4-128_N8000_ab7_lamb2.5}(c) inset], for which the
effective conductance $G$ between nodes $A$ and $B$ satisfies
\begin{equation}
\frac{1}{G}=\frac{1}{G_A}+\frac{1}{G_{tb}}+\frac{1}{G_B},
\label{e4a}
\end{equation}
where $1/G_{tb}$ is the resistance of the ``transport backbone'' while
$1/G_A$ (and $1/G_B$) are the resistances of the set of bonds near node
$A$ (and node $B$) not belonging to the ``transport backbone''. It is
plausible that $G_A$ is linear in $k_A$, so we can write
$G_A=ck_A$. Since node $B$ is equivalent to node $A$, we expect
$G_B=ck_B$. Hence
\begin{equation}
G= \frac{1}{1/ck_A +1/ck_B+1/G_{tb}}
=k_B\frac{ck_A/k_B}{1+k_A/k_B+ck_A/G_{tb}},
\label{Gh}
\end{equation}
so the scaling function defined in Eq.~(\ref{G_max_scaled}) is
\begin{equation}
f(x)=\frac{cx}{1+x+ck_A/G_{tb}}\approx\frac{cx}{1+x}.
\label{e4b}
\end{equation}
The second equality follows if there are many parallel paths on the
``transport backbone'' so that $1/G_{tb}\ll ck_A$ \cite{text1}.  The
prediction (\ref{e4b}) is plotted in
Fig.~\ref{PGka750_kb4-128_N8000_ab7_lamb2.5}(c) and the agreement with
the simulations supports the approximate validity of the transport
backbone picture of conductance in complex networks.

The agreement of (\ref{e4b}) with simulations has a striking
implication: the conductance of a scale-free network depends on only one
parameter $c$. Further, since the distribution of
Fig.~\ref{PGka750_kb4-128_N8000_ab7_lamb2.5}(a) is sharply peaked, a
single measurement of $G$ for any values of the degrees $k_A$ and $k_B$
of the entrance and exit nodes suffices to determine $G^\ast$, which
then determines $c$ and hence through Eq.~(\ref{e4b}) the conductance 
for all values of $k_A$ and $k_B$.

\begin{figure}[t]
\epsfig{file=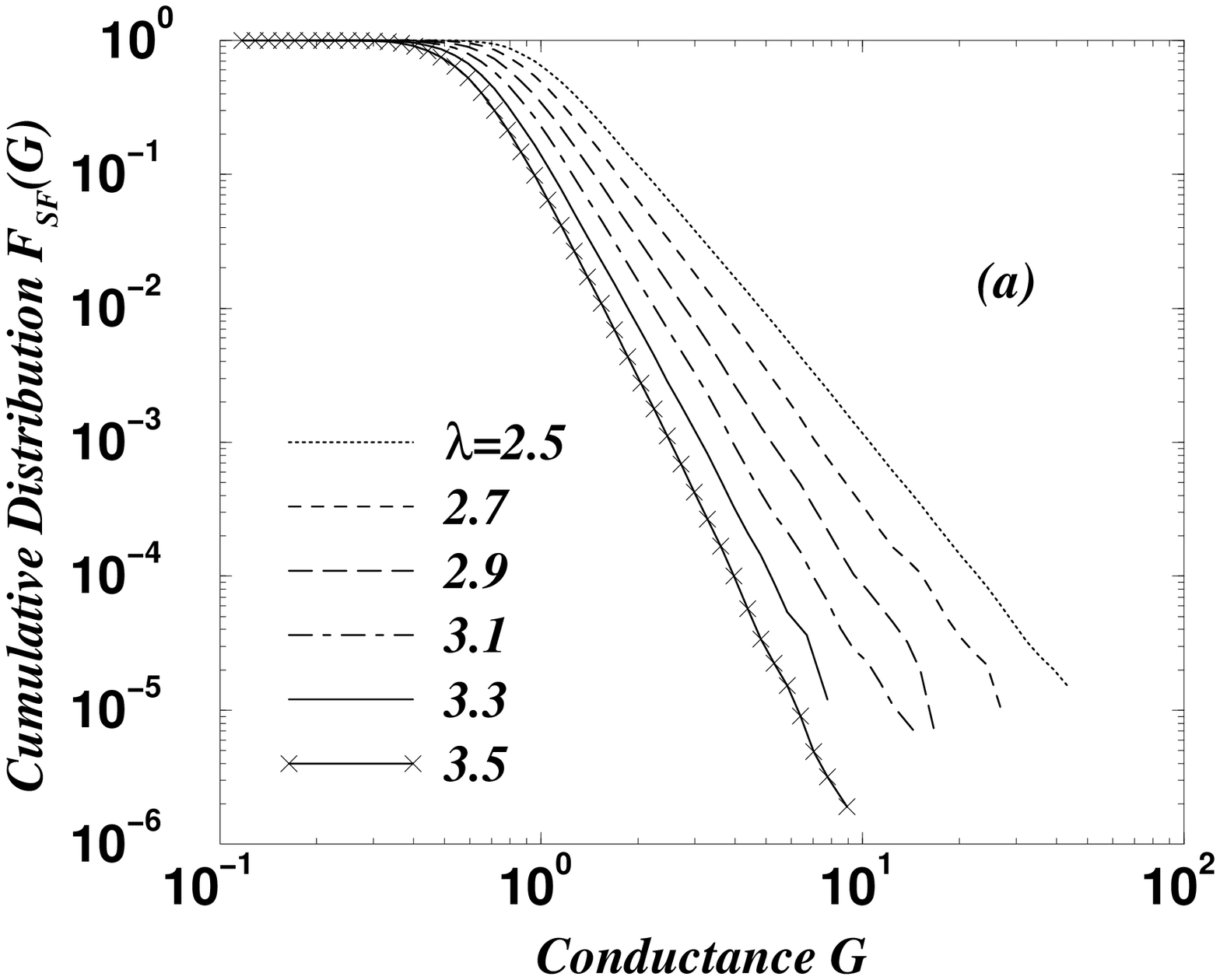,width=3in,height=2.5in}
\epsfig{file=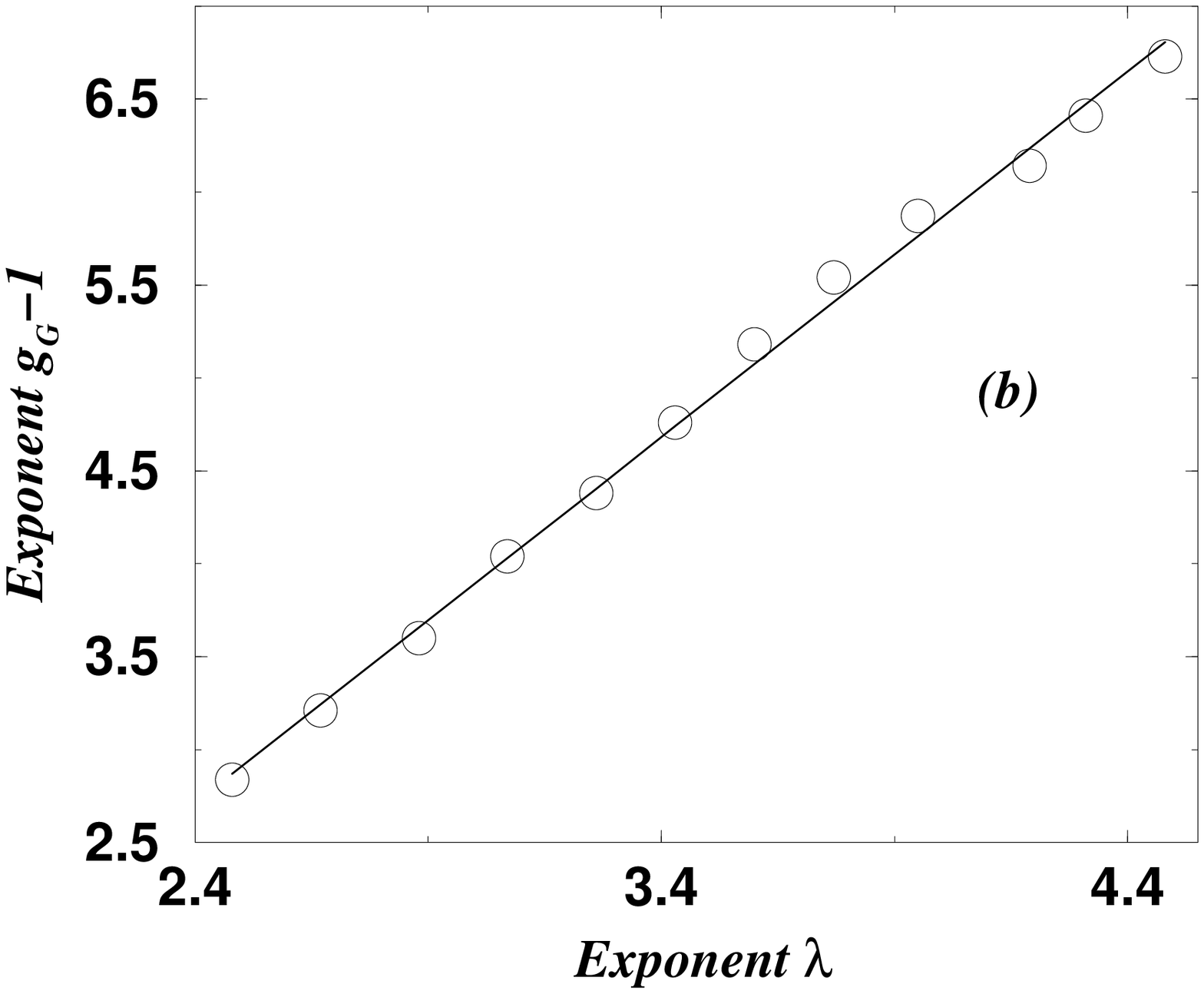,width=3in,height=2.5in}
\caption{(a) Simulation results for the cumulative distribution $F_{\rm
     SF}(G)$ for $\lambda$ between 2.5 and 3.5, consistent with the
     power law $F_{\rm SF}\sim G^{-(g_G-1)}$ (cf. Eq.~(\ref{PhiG})),
     showing the progressive change of the slope $g_G-1$. (b) The
     exponent $g_G-1$ from simulations (circles) with $2.5<\lambda<4.5$;
     shown also is a least
     square fit $g_G -1=(1.97\pm 0.04)\lambda -(2.01\pm 0.13)$,
     consistent with the predicted expression $g_G-1=2\lambda -2$
     [cf. Eq.~(\ref{g_G})].}
\label{FG_G_lamb2.5-3.5_8000}
\end{figure}

Within this ``transport backbone'' picture, we can analytically 
calculate $F_{\rm SF}(G)$.  Using Eq.~(\ref{G_max_scaled}), 
and the fact that $\Phi_{\rm
SF}(G|k_A,k_B)$ is narrow, yields \cite{VanKampen}
\begin{equation}
\Phi_{\rm SF}(G)\sim \int P(k_B)dk_B
\int P(k_A)dk_A \delta\left[k_B f\left(\frac{k_A}{k_B}\right)-G\right],
\label{Gkakbh}
\end{equation}
where $\delta(x)$ is the Dirac delta function. 
Performing the integration of Eq.~(\ref{Gkakbh}) using (\ref{e4b}), we obtain 
\begin{equation}
\Phi_{\rm SF}(G)\sim G^{-g_G} \qquad [G<G_{\rm max}],
\label{PhiG}
\end{equation}
where
\begin{equation}
g_G=2\lambda -1.
\label{g_G}
\end{equation}
Hence, for $F_{\rm SF}(G)$, we have 
$F_{\rm SF}(G)\sim G^{-2\lambda -2}$.
To test this prediction, we perform simulations for scale-free networks
and calculate the values of $g_G-1$ from the slope of a log-log plot of
the cumulative distribution $F_{\rm SF}(G)$.  From
Fig.~\ref{FG_G_lamb2.5-3.5_8000}(b) we find that
\begin{equation}
g_G -1=(1.97\pm 0.04)\lambda -(2.01\pm 0.13).
\end{equation}
Thus, the measured slopes are consistent with the theoretical value
predicted by Eq.~(\ref{g_G}) \cite{fn_lambda}.

In summary, we find a power-law tail for the distribution of conductance
for scale-free networks and relate the tail exponent $g_G$ to the
exponent $\lambda$ of the distribution $P(k)$.  Our work is consistent
with a simple physical picture of how transport takes place in complex
networks.

\noindent
We thank the Office of Naval Research, the Israel Science Foundation,
and the Israeli Center for Complexity Science for financial support, and
L. Braunstein, S. Carmi, R. Cohen, E. Perlsman, G. Paul, S. Sreenivasan,
T. Tanizawa, and Z. Wu for discussions.

\eject

\eject

\begin{thebibliography}{99}
\bibitem{netwcomment} In principle, a node can have a degree up to
$N-1$, connecting to all other nodes of the network.  The results
presented here corresponds to networks with the upper cutoff $k_{\rm
max}=k_{\rm min}N^{1/(\lambda-1)}$ imposed.  We also studied networks
for which this cutoff is not imposed, and found no significant
differences in the pdf $\Phi_{\rm SF}(G)$.
\bibitem{dyn-topology} The dynamical properties that we study here are
related to transport on networks and differ from those which treat the 
network topology itself as evolving in time \cite{scale-Barabasi,dyn-network}.
\bibitem{fn1} We performed simulations 
in which the random copies are generated and then randomly matched, and also
simulated networks where the node copies were paired in order of degree
from highest to lowest and obtained similar results.
\bibitem{Gavg} Since $g_G>2$ for $\lambda>2$, it follows that although
$\bar{G}$ increases with $N$, it finally converges to a constant. Only
for $g_G<2$, the cutoff $G_{\rm max}$ in the tail of $\Phi_{\rm SF}(G)$ 
controls $\bar{G}$, making $\bar{G}$ an increasing unbounded function of $N$.
\bibitem{text1}
Flux starts at node $A$, being controlled by the conductance of the
bonds in the vicinity of node $A$. This flux passes into the ``transport
backbone'', which is comprised of many parallel paths and hence has a
high conductance.  Finally, flux ends at node $B$,
being controlled by the conductance of the bonds in the vicinity of node $B$.
This is similar to the traffic around a major
freeway.  Most of the limitations to transport occur in getting to the
freeway (``node $A$'') and then after leaving it (``node $B$''), 
but flow occurs easily on the freeway (``transport backbone'').
\bibitem{fn_lambda} The $\lambda$ values explored here are limited by 
computer time considerations. In principle, we are unaware of any theoretical
reason that would limit the validity of our results to any particular range
of values of $\lambda$ (above 2).
\end{thebibliography}
\end{document}